\documentstyle[twocolumn,prb,aps,floats,epsfig]{revtex}
\begin{document}
\twocolumn[\hsize\textwidth\columnwidth\hsize\csname 
@twocolumnfalse\endcsname
\title{A model of transport nonuniversality in
thick-film resistors} 
\author{C. Grimaldi}
\address{Institut de Production et Robotique, LPM,
EPFL,
CH-1015 Lausanne, Switzerland}
\author{T. Maeder}
\address{Institut de Production et Robotique, LPM,
EPFL,
CH-1015 Lausanne, Switzerland}
\address{Sensile Technologies SA, PSE, CH-1015 Lausanne, Switzerland}
\author{P. Ryser}
\address{Institut de Production et Robotique, LPM,
EPFL,
CH-1015 Lausanne, Switzerland}
\author{S. Str\"assler} 
\address{Institut de Production et Robotique, LPM,
EPFL,
CH-1015 Lausanne, Switzerland}
\address{Sensile Technologies SA, PSE, CH-1015 Lausanne, Switzerland}
\maketitle

%\centerline \\

\begin{abstract}
We propose a model of transport in thick-film resistors which naturally
explains the observed nonuniversal values of the conductance exponent $t$
extracted in the vicinity of the percolation transition. 
Essential ingredients of the model are the segregated microstructure typical
of thick-film resistors and tunneling between the conducting grains.
Nonuniversality sets in as consequence of wide distribution of interparticle
tunneling distances.

\end{abstract}
\vskip 2pc ]

\newpage

\begin{figure}
\protect
\centerline{\epsfig{figure=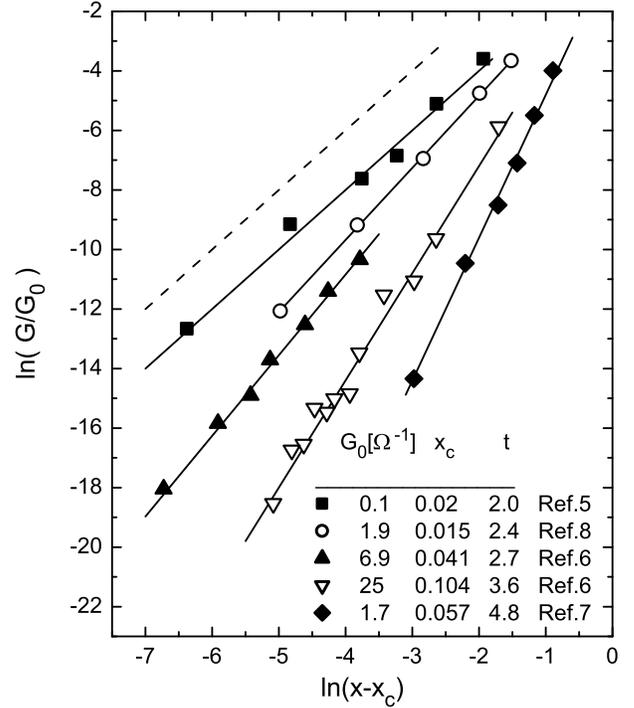,width=20pc,clip=}}
\caption{Measured conductances on different RuO$_2$ 
(Refs.\protect\onlinecite{carcia2,tambo}) and Bi$_2$Ru$_2$O$_7$ (Refs.\protect\onlinecite{carcia1,listki})
TFRs. Solid lines are fits to Eq.(\ref{eq1}) with fitting values reported in the
inset. dashed line denotes a power-law with exponent $t=2$.}
\label{fig1}
\end{figure}

%\narrowtext
\centerline \\
Thick-film resistors (TFRs) are glass-conductor
composites based on RuO$_2$ (but also Bi$_2$Ru$_2$O$_7$,
Pb$_2$Ru$_2$O$_6$, and IrO$_2$) grains mixed and fired with glass powders.\cite{prude1}
Besides the widespread use of TFRs in
pressure and force sensor applications,\cite{white} their transport properties are
of great interest also for basic research. 
The percolating nature of transport in TFRs
has been reported since long time and now 
it is well documented.\cite{pike,dejeu,carcia1,carcia2,listki,tambo,kusy}
As shown in Fig.\ref{fig1} where we reports a selection of previously published data
on different TFRs,\cite{carcia1,carcia2,listki,tambo}
the conducting phase concentration $x$ dependence of the conductance
$G$ of TFRs follows a percolating-like power-law equation of the form:
\begin{equation}
\label{eq1}
G=G_0(x-x_c)^t,
\end{equation}
where $G_0$ is a prefactor, $x_c$ is the critical concentration below which 
$G$ vanishes and $t$ is the transport critical exponent.\cite{kirk,stauffer} 
The values of $G_0$, $x_c$, and $t$ which best fit the experimental data
are reported in the inset of Fig.\ref{fig1}. 

According to the standard theory of transport in isotropic
percolating systems,\cite{stauffer}
$G_0$ and $x_c$  depend on microscopic details such as the microstructure
and the mean value of the junction resistances connecting two 
neighbouring conducting sites, while, unless the microscopic resistances have a 
diverging distribution function (see below),
the critical exponent $t$ is {\it universal}, {\it i. e.}, it depends only upon
the lattice dimensionality $D$. For $D=3$, random resistor network calculations
predict $t=t_0\simeq 2.0$,\cite{clerc} in agreement with various granular metal
systems,\cite{abeles,lee} or other disordered compounds.\cite{putten}

As it is clear from Fig.\ref{fig1}, TFRs have values of $t$ ranging from its
universal limit $t\simeq 2.0$ (filled squares, Ref.\onlinecite{carcia1})
up to very high values like $t\sim 5.0$ (filled diamonds, Ref.\onlinecite{listki})
or even higher.\cite{pike}
Despite of their clear percolating behavior, TFRs do not fulfil
therefore the hypothesis of universality common to many systems and instead belong
to a different, quite vast, class of materials which display 
nonuniversal transport behavior, that is a regime where the transport exponent $t$ 
depends on microscopic details (microstructure etc.).
Typical examples of nonuniversal systems are carbon-black--polymer composites,\cite{balb}
and materials constituted by insulating regions embedded in a 
continuous conducting phase.\cite{lee,wu} 

Despite that TFRs have been historically among
the first materials for which transport nonuniversality has been reported,\cite{pike} 
the microscopic origin of their universality breakdown has not been specifically
addressed so far. 
In this letter we show that the cross-over between universality and 
nonuniversality reported in Fig.\ref{fig1} can be explained 
within a single model whose basic features are the peculiar microstructure of TFRs
and the tunneling processes between conducting grains.

Before describing our model for TFRs, let us first recall the mathematical
requisites for universality breakdown in random resistor networks. Consider a regular 
lattice of sites and assign to each neighbouring couple of sites a bond which
has finite conductance $g$ with probability $p$ and zero conductance
with probability $1-p$.
The resulting conductance distribution function is then:
\begin{equation}
\label{distri1}
\rho(g)=p h(g)+(1-p)\delta (g),
\end{equation}
where $\delta(g)$ is the Dirac delta-function and $h(g)$ is the distribution
function of the finite bond conductances. For well behaved distribution functions
$h(g)$, conductivity is universal and follows Eq.(\ref{eq1}) with $t=t_0$. 
Instead, as first shown by Kogut and Straley,\cite{straley} 
if $h(g)$ has a power law divergence for small $g$
of the form:
\begin{equation}
\label{distri2}
\lim_{g\rightarrow 0} h(g) \propto g^{-\alpha},
\end{equation}
and $\alpha$ is larger than a critical value $\alpha_c$, then transport is no longer 
universal. Renormalization group analysis predicts in fact that
\begin{equation}
\label{nonuni}
t=\left\{
\begin{array}{ll}
t_0 & \mbox{if} \hspace{3mm}\nu+1/(1-\alpha)< t_0 \\
\nu+1/(1-\alpha) &  \mbox{if}\hspace{3mm} \nu+1/(1-\alpha)> t_0
\end{array}
\right.,
\end{equation}
where $\nu\simeq 0.88$ is the
correlation-length exponent for a three dimensional lattice.\cite{machta} 
By using $t_0\simeq 2.0$ we obtain therefore $\alpha_c\simeq 0.107$.
Equations (\ref{distri2}) and (\ref{nonuni}) have been shown to arise from
a system of insulating spheres embedded in a continuous 
conducting material (swiss-cheese model),\cite{halperin} and from a 
tunneling-percolation model
with highly fluctuating tunneling distances.\cite{balb}
Here we show that Eq.(\ref{distri2}) [and consequently Eq.(\ref{nonuni})]
arises naturally from a simple representation of TFRs in terms of 
their microstructure and elemental transport processes.

Let us start by considering the highly non-homogeneous
microstructure typical of TFRs. These systems are constituted by a mixture
of large glassy particles (typically with size $L$ of order $1$-$3$ $\mu$m) and small
conducting grains of size $\Phi$ varying between $\sim 10$ nm up to $\sim 200$ nm.
In this situation, the small metallic grains tend to occupy the narrow 
regions between the much larger insulating zones leading to a filamentary
distribution of the conducting phase. A classical model to describe
such a segregation effect was proposed already in the 1970's by Pike.\cite{pike} 
This model treats the
glassy particles as cubes of size $L\gg \Phi$ whose edges are occupied by chains
of adjacent metallic spheres of diameter $\Phi$. Such chains define channels (bonds)
which form a cubic lattice spanning the whole sample. Let us assume for the moment
that a bond has probability $p$ of being occupied by a fixed number $n+1$ of spheres
and probability $1-p$ of being empty. To each couple of adjacent spheres we assign
an inter-sphere conductance $\sigma_i$ ($i=1,\cdots ,n$).
A random resistor network can be therefore defined as in Eq.(\ref{distri1})
where $h_n(g)$ is the distribution function of the total channel conductance $g$ 
of $n$ conductances $\sigma_i$ in series:
\begin{equation}
\label{bond1}
g^{-1}=\sum_{i=1}^{n}\frac{1}{\sigma_i}.
\end{equation}
The high values of piezoresistance ({\it i.e.}, the strain sensitivity of transport)
typical of TFRs,\cite{prude1} and the low temperature dependence of transport 
strongly indicate that the main contribution to
the overall resistance stems from tunneling processes between neighbouring 
metallic grains.
Hence, if the centers of two neighbouring metallic spheres are separated by a 
distance $r$, then the intergrain tunneling
conductance $\sigma$ is approximatively of the form:
\begin{equation}
\label{tunnel1}
\sigma=\sigma(r)\equiv\sigma_0 e^{-2(r-\Phi)/\xi},
\end{equation}
where $\sigma_0$ is a constant which we set equal to the unity, 
$\xi \propto 1/\sqrt{V}$ is the tunneling factor and
$V$ is the intergrain barrier potential. 
Let us make the quite general assumption that the centers of the spheres
are set randomly along the channel, so that the distances $r$ change according
to the distribution function $P(r)$ for a set of impenetrable spheres arranged randomly
in a quasi one dimensional channel. By following Ref.\cite{torquato}, $P(r)$ can be 
calculated exactly and its explicit expression is:
\begin{equation}
\label{tran}
P(r)=\frac{1}{a_n-\Phi}e^{-(r-\Phi)/(a_n-\Phi)}\Theta(r-\Phi) 
\end{equation}
where $a_n=(1+L/n\Phi)\Phi/2$ is the mean inter-sphere distance and 
$\Theta$ is the step function. By combining Eq.(\ref{tunnel1}) with Eq.(\ref{tran})
the distribution $f(\sigma)$ of the inter-sphere conductances is then:
\begin{equation}
\label{distri4}
f(\sigma)=\int_0^{\infty} \!dr\, P(r)\,\delta[\sigma-\sigma(r)]
=(1-\alpha_n)\sigma^{-\alpha_n},
\end{equation}
where
\begin{equation}
\label{alfa}
\alpha_n=1-\frac{\xi/2}{a_n-\Phi}.
\end{equation}
To obtain the distribution function $h_n(g)$ of the occupied channels, we first note
that Eq.(\ref{bond1}) implyies that $g$ is dominated by the minimum inter-sphere
conductance $\sigma_{\rm min}$ among the set of $n$ conductances in series. Hence
the small-$g$ limit of $h_n(g)$ is just the distribution function $\tilde{f}$ 
of $\sigma_{\rm min}$:
\begin{equation}
\label{min}
\tilde{f}(\sigma_{\rm min})=
n f(\sigma_{\rm min})\left[1-\int_{\sigma_{\rm min}}^1 d\sigma_{\rm min}
f(\sigma_{\rm min})\right]^{n-1},
\end{equation}
which, from Eq.(\ref{distri4}) and by setting $g\simeq \sigma_{\rm min}$, leads to:
\begin{equation}
\label{distri5}
\lim_{g\rightarrow 0} h_n(g)\simeq n(1-\alpha_n)g^{-\alpha_n}.
\end{equation}

The conducting bond distribution function behaves therefore as Eq.(\ref{distri2})
so that for $\alpha_n>\alpha_c\simeq 0.107$ transport universality breaks down and
$t>t_0$. For $L=1$ $\mu$m, $\Phi=10$ nm, and $\xi=1$ nm this is achieved already for
$n<90$, {\it i. e.}, slightly less then the maximum number $L/\Phi=100$ 
of spheres which can be accommodated inside a channel. 

Our model of universality breakdown in TFRs can be readily generalized to describe
more realistic situations. For example, the number $n$
of spheres inside the occupied channels can vary according to a given 
distribution. 
It is also straightforward to rewrite Eq.(\ref{tran}) in order 
to describe cases in which
the diameter $\Phi$ of the spheres is not fixed,\cite{lu} or to let the size of the
insulating grains to change by assigning a distribution function for $L$.
It is then possible to have different scenarios and, more importantly, to 
obtain a crossover from transport universality ($t=t_0$) to nonuniversality 
($t>t_0$) within the same framework. This reminds the experimental situation
reported for TFRs and summarized in Fig.\ref{fig1}.

Let us now comment on the capability of other existing theories to describe transport
universality breakdown in TFRs. At a first glance, the swiss-cheese model
of Ref.\onlinecite{halperin} is a natural candidate since the large values
of $L/\Phi$ typical of many TFRs may lead to an effective continuous conducting
phase filling the voids between the large glassy grains. However, there are examples
in which nonuniversality has been reported for TFRs with $L/\Phi$ only of
order $\sim 5-10$,\cite{carcia1} a value probably too small to be compatible 
with the swiss-cheese picture.
Even more problematic are the cases for
which $t=t_0\simeq 2.0$ has been measured for TFRs with $L/\Phi\sim 100$
(see for example Fig.\ref{fig1}), while the swiss-cheese model would have 
predicted $t>t_0$.
Regarding instead the model proposed by Balberg,\cite{balb} for 
which transport is dominated by 
random tunneling processes in a percolating network, it is important to point
out that it was defined by using a phenomenological distribution function for
the nearest-neighbour particle distances very similar to our Eq.(\ref{tran}).
Balberg argured that such form of $P(r)$ is a reasonable compromise between
the distribution function of spheres 
randomly placed in three dimensions,\cite{torquato}
and the effect of interactions between the conducting and insulating phases.
Instead we have shown that Eq.(\ref{tran}), 
and consequently the power-law divergence of $h(g)$, is a straightforward outcome
of the quasi one-dimensional geometry of the conducting channels in 
the segregation model of TFRs.  Despite that our model has been formulated
specifically for TRFs, nevertheless it could be applied also to other
segregated disordered compounds for which tunneling is the main mechanism
of transport and nonuniversality has been reported.\cite{wu}

In conclusion, we have proposed a simple tunneling-percolation model capable of
describing the observed transport universality breakdown in TFRs. 
Essential ingredients of the theory are the segregated structure, modelled by 
quasi-one dimensional channels occupied randomly by the conducting particles, and 
intergrain tunneling taking place within the channels. 

This work is part of TOPNANO 21 project n.5557.2.

\end{document}